\documentclass[onecolumn, draftcls]{IEEEtran}

\usepackage{amsmath}
\usepackage{cases}
\usepackage{amsfonts}
\usepackage{amssymb}
\usepackage{boxedminipage}
\usepackage{graphicx}
\usepackage[usenames]{color}
\usepackage{array,color}
\usepackage{colortbl}
\usepackage{bm}
\usepackage{supertabular}

\begin{document}
 \baselineskip=12pt
\title{Every-User Delay Guarantee for Wireless Multiple Access Systems}
\author{Chuang~Zhang$^\ast$, ~Pingyi~Fan$^\ast$, ~Ke~Xiong$^\S$, ~Yunquan~Dong$^\ast$\\  
$^\ast$Department of Electronic Engineering, Tsinghua University, Beijing, P.R. China\\
$^\S$School of Computer and Information Technology, Beijing Jiaotong University, Beijing, P.R. China\\
E-mail:~\{zhangchuang11,dongyq08\}@mails.tsinghua.edu.cn,~fpy@tsinghua.edu.cn,~kxiong@bjtu.edu.cn}
\maketitle

\begin{abstract}
The quality of service (QoS) requirements are usually different from user to user in a multiaccess system, and it is necessary to take the different requirements into account when allocating the shared resources of the system. In this paper, we consider one QoS criterion--delay in a multiaccess system, and we combine information theory and queueing theory in an attempt to analyze whether a multiaccess system can meet the different delay requirements of users. For users with the same transmission power, we prove that only $N$ inequalities are necessary for the checking, and for users with different transmission powers, we provide a polynomial-time algorithm for such a decision. In cases where the system cannot satisfy the delay requirements of all users, we prove that as long as the sum power is larger than a threshold, there is always an approach to adjust the transmission power of each user to make the system delay feasible if power reallocation is available.
\end{abstract}


\section{Introduction}

Most of today's wireless communication systems are multiaccess systems in the uplink. Typically, a set of users with different quality of service (QoS) requirements compete for the usage of common resources like bandwidth, power etc., besides, even the same user may vary his requirement for the system according to the kind of service he is receiving. To better allocate the resources in a multiaccess system, it is necessary to take the different QoS requirements into account, since in this way, the system can avoid putting too much effort on over-serving one user and save the energy to meet the requirements of many other users. In this preliminary work, we focus our attention on satisfying one QoS criterion--delay requirements, and we try to decide whether a multiaccess system can meet the different delay requirements of all users.

Some previous works have discussed the delay problem in a multiaccess system \cite{yeh2002doramac}-\cite{mandal2012drpma}. However, due to its difficulty, most of these works only discussed minimizing the average delay among all users in multiaccess systems and didn't consider guaranteeing delay requirement for each specific user. For instance, E. Yeh \cite{yeh2002doramac} proposed a \emph{Longer-Queue-Higher-Rate} (LQHR) method to minimize the average delay of a  multiaccess system with symmetric arrival rates. N. Ehsan  et al. \cite{ehsan2001dotpwmc} proved that the delay-optimal rate allocation policy for a multiaccess system with asymmetric arrival rates is of threshold type. Jing Yang et al. \cite{jing2010dmtfapcmac} analyzed the delay minimization problem using a multi-dimensional Markov chain, and showed that the optimal transmission policy is also of a threshold structure.

In this paper, we investigate the delay-guaranteed transmission for multiaccess systems. Different from existing works which aimed to minimize the statistical average delay among all multiaccess users, we consider providing service for users with different delay requirements. Therefore, our focus is not on minimizing the average delay, but on determining whether a system can satisfy the different needs of all users, and in cases where the system is not able to meet the requirements of all users, we seek to find some ways such as power reallocation to make the system delay feasible. In addition, the number of users in our discussion is not constrained to two, in fact, the advantage of the algorithm we use is more obvious for systems with large number of users.

The rest of the paper is organized as follows, the system model and the multiaccess channel capacity region are introduced in Section \ref{sec_model} and Section \ref{sec_macr} respectively. In Section \ref{sec_wmamdr}, we discuss how to determine whether the multiaccess system can meet the different delay requirements, and apply a polynomial-time algorithm for the general case. In Section \ref{sec_poweralloc} we present an explicit threshold of sum power to meet the delay requirements and provide a power allocation method when the sum power is larger than the threshold. Finally, the conclusion is given in Section \ref{sec_conclusion}.

\section{System Model}\label{sec_model}

\begin{figure}
  \centering
  \includegraphics[width=0.4\textwidth]{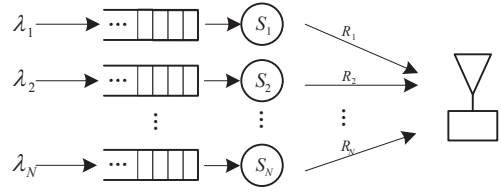}\\
  \caption{A multiaccess system with $N$ users. }\label{fig_MAmodel}
\end{figure}

We consider a multiaccess system, as shown in Fig. \ref{fig_MAmodel}. In this system, $N$ users communicate to the same receiver with transmission power $P_1,P_2,\ldots,P_N,$ respectively. The packet arrival process of each user is Poisson, with average positive arrival rate $\lambda_1,\lambda_2,\ldots,\lambda_N,$ respectively. Besides, we assume that each packet has the same fixed length $L$ (assuming $L=1$ in our later discussions). If each user has a delay requirement $\tau_i, i\in \{1,\ldots,N\}$, we want to know whether this multiaccess system can meet the delay requirements of all users.

This problem can be treated as a cross layer design problem and we follow the line of Gallager \cite{telatar1995qtitm} which combined queueing theory and information theory for the multiaccess system. Specifically, we divide this system into two layers, the physical layer and the medium access layer. The physical layer determines the service rate of the system, while the medium access layer is related to the delay and can be modeled by a queueing process. In addition, the buffer of each user is assumed to be of infinite length, therefore, the queueing model of each user is M/G/1. We consider allocating service rate for each user in the system. It is know that for a multiaccess additive Gaussian noise channel, the capacity region is determined by the transmission power of each user, we can first decide the required service rate of each user based on the delay requirement, and then determine whether the service rate vector lies in the capacity region of the multiaccess system. If it does, the multiaccess system can satisfy the delay requirement of each user. If not, the multiaccess system cannot meet the delay requirement. And for the case where the delay requirement is not satisfied, we find the necessary condition to meet the minimum delay requirement of each user which can be characterized as a threshold. Furthermore, we propose a power allocation policy for feasible sum power.

In order to fit the rate vector into the multiaccess capacity region, it is expected that the required service rate of each user is as small as possible. It can be proved that the strategy of allocating a fixed rate to each user results in the smallest average service rate. This is emphasized by the following lemma.

\emph{\textbf{Lemma 1:}} \emph{For a M/G/1 queueing system, if the average arrival rate is $\lambda$, the average sojourn time is $\tau$, then the minimum average service rate is obtained when the service process is a deterministic process, i.e., when the M/G/1 queueing system reduces to M/D/1}.
\begin{proof}
From the formula of Pollaczek-Khintchine \cite{wolff1989smattq}, which is the average queue length of M/G/1 and Little's Law, we get the average sojourn time of a M/G/1 queueing system,
\begin{equation}\label{avert}
  \tau=\frac{1}{R}+\frac{\frac{\lambda}{R^2}(1+c_{b}^{2})}{2(1-\frac{\lambda}{R})},
\end{equation}
where $R$ is the average service rate and $c_{b}^{2}$ is the coefficient of variance of the random variable service time.  Then the average service rate can be expressed as
\begin{equation}\label{equ_asr}
  R=\frac{(\lambda\tau+1)+\sqrt{\lambda^2\tau^2+2\lambda\tau c^2_b+1}}{2\tau}.
\end{equation}
It can be seen from Eqn. (\ref{equ_asr}) that the minimum average service rate is obtained when the coefficient of variance of service time $c_b$ equals $0$, i.e., when the service process is a deterministic process.
\end{proof}

Lemma 1 indicates that if we cannot find a fixed point in the capacity region to meet the delay requirements of all users, we cannot meet the requirements by allocating the rate vector in the capacity region according to some probability either. So to decide whether a multiaccess system can meet the delay requirements of all users, we only need to check the existence of such fixed point in the capacity region. 

It needs to be noted that allocation of a fixed point in the capacity region or according to some distribution applies to situations where the queue state information of all users is not available to the common receiver, in cases where the receiver can obtain the information, it can allocate the rate dynamically based on the queue state information, and thus improve the delay performance. For instance, considering a situation where the load is light for each user, and only one queue is not empty in each time slot, the system can easily provide each user with the highest service rate. However, if each user is heavily loaded, to the other extreme, when each queue is infinitely backlogged, dynamic allocation of the rates wouldn't improve the delay performance since when the system goes stable, dynamic allocation is equal to rate allocation according to some distribution. So our following discussions are based on the assumption that the queue state information is not available or when the system is heavily loaded.

\section{The multiaccess Capacity Region} \label{sec_macr}

Consider a $N$-user multiaccess additive Gaussian noise channel with noise density $N_0/2$ and two-sided bandwidth $2W$, if the transmission power of each user is $P_1,P_2,\ldots,P_N$ respectively, for $\forall S\subseteq\{1,\ldots,N\}$, the capacity region of this multiaccess system is
\begin{equation}\label{equ_capregion}
\begin{split}
  C_g(\bm{P})=\{\bm{R}:\bm{R}(S)\leq W\log(1+\tfrac{\sum_{i\in S}P_i}{N_0W}), \\
\end{split}
\end{equation}
where $\bm{R}(S)=\sum_{i\in S} R_i$.

It can be seen that in order to determine whether a rate vector lies in the capacity region, we need to check $2^N-1$ inequalities. This checking process is a huge burden when the number of users becomes large, so we need to devise algorithms to reduce the computational complexity. We discuss this problem in two different cases. In the first case, we assume that the transmission powers of all users are the same, and we find that only $N$ inequalities are necessary for the checking. For the second case where the transmission power of each user is different, we provide a polynomial-time algorithm based on submodular function minimization.

\section{Whether the multiaccess system can meet the delay requirements}\label{sec_wmamdr}

According to the above discussions, we can divide the process of deciding whether the multiaccess system can meet the delay requirements into two separate steps. In the first step, based on the M/D/1 queueing model, we calculate the required service rate vector. In the second step, we determine whether the required service rate vector lies in the capacity region of the multiaccess system.

The required service rate vector can be easily calculated as
\begin{flalign}\label{equ_rsrv}
\bm{R}&=\{ (R_i)| R_i=\tfrac{(\lambda_i\tau_i+1)+\sqrt{\lambda_i^2\tau_i^2+1}}{2\tau_i},i\in\{1,\ldots,N\}\},
\end{flalign}
where $N$ is the number of users in the system. Then we need to determine whether this rate vector $\bm{R}$ lies in the multiaccess capacity region.

\subsection{Users with the same transmission power}

We first consider a special case where the transmission powers of all users are the same, in this case, the checking process can be simplified significantly. This is illustrated in the following proposition.

\textbf{\emph{Propositon 1:}} \emph{For users with the same transmission power, we only need to check $N$ inequalities to decide whether the multiaccess system can satisfy the delay requirements.}
\begin{proof}
Let $\Pi=(\pi(1),\pi(2),\ldots,\pi(N))$ be a sequence which satisfies $R_{\pi(1)}\geq R_{\pi(2)}\geq \ldots \geq R_{\pi(N)}$, then we only need to check the following inequalities to decide whether the rate vector lies in the capacity region
$$
\left\{
\begin{aligned}
  R_{\pi(1)} &\leq& W\log(1+\tfrac{P}{N_0W}) \nonumber\\
  R_{\pi(1)}+R_{\pi(2)} &\leq& W\log(1+\tfrac{2P}{N_0W}) \nonumber\\
  &\vdots& \nonumber\\
  R_{\pi(1)}+R_{\pi(2)}+\ldots+R_{\pi(N)} &\leq& W\log(1+\tfrac{NP}{N_0W})
\end{aligned}
\right. .
$$

It is easy to see that if these $N$ inequalities hold, all of the other inequalities will also hold. For instance, we know that
$$
\sum\nolimits_{i\in S}R_i\leq \sum\nolimits_{i=1}^{|S|}R_{\pi(i)}\leq W\log(1+\tfrac{|S|P}{N_0W}),
$$
for the user in $S\subseteq\{1,2,\ldots,N\}$,where $|S|$ denotes the cardinality of the set $S$. Thus, checking the $N$ inequalities above is enough to determine whether the rate vector lies in the capacity region.
\end{proof}

\subsection{Users with different transmission powers}\label{sec_conclusion}

If the transmission power of each user is different, normally we need to check  $2^{N}-1$ inequalities to decide whether the rate vector lies in the capacity region, the computational complexity becomes very high when the number of users is very large, so it is necessary to find an algorithm to reduce the computational complexity.

In \cite{tse1998multfc}, it was noted that the capacity region of the multiaccess system has a polymatroid structure, and we utilized this structure in searching for a polynomial-time algorithm.

First, we give an introduction to polymatroid. In the following, we use $E$ to denote the set $\{1,2,\ldots,N\}$, and use $\bm{v}(S)$ to denote $\sum_{i\in S}v_i$ for the vector $\bm{v} \in R^E$.

\emph{\textbf{Definition 1:}} The set $P(f)=\{\bm{v}|\bm{v}\in R^{E}, \forall S\subseteq E:\bm{v}(S)\leq f(S) \}$ is a polymatroid if the function $f: 2^{E}\longrightarrow R_{+}$ satisfies

1) $f(\emptyset)=0$, (normalized),

2) $f(S)\leq f(T), \forall S\subseteq T$, (nondecreasing),

3) $f(S)+f(T)\geq f(S\cup T)+f(S\cap T)$, (submodular).

It is easy to see that the capacity region of a multiaccess system (\ref{equ_capregion}) is a polymatroid.

Specifically, considering the property of submodularity of a polymatroid structure, we apply the algorithm proposed in \cite{sat2001acspa} to minimize the submodular function in our problem.

%

It can be seen that the function
$g(S)=W\log(1+\tfrac{\sum\nolimits_{i\in S}P_i}{N_0W})$
is a submodular function. In fact, the function
$f(S)=g(S)-\sum\nolimits_{i\in S}R_i$ is also submodular.

The problem of testing whether a rate vector lies in the capacity region is equivalent to finding a minimum for the submodular function $f(S)$ defined above, if $\min\{f(S)\}\geq 0$, then $\forall S\subseteq E$, we have $\bm{R}(S)\leq W\log(1+\frac{\sum_{i\in S}P_i}{N_0W})$, and the rate vector lies in the capacity region. On the contrary, if $\min\{f(S)\}<0$, then there exists at least one $S\subseteq E$ which renders $\bm{R}(S)> W\log(1+\frac{\sum_{i\in S}P_i}{N_0W})$. In this case, the rate vector does not lie in the capacity region. Thus, the membership testing problem is transformed into a submodular function minimization problem.

In the literature, Gr\"{o}tschel et al. \cite{gro1981temac} proposed a polynomial-time algorithm for submodular function minimization using ellipsoid method, however, the ellipsoid method itself requires much computation, thus this algorithm is not efficient in practice. In \cite{whc1984tmimp}, Cunningham proposed a polynomial-time algorithm for testing membership in matroid polyhedra, the algorithm used an augmenting path approach and can efficiently determine whether a nonnegative real vector is in the convex hull of independent sets of a matroid, but it can not be applied to a polymatroid structure. Inspired by the augmenting path approach, Satoru Iwata et al. \cite{sat2001acspa} proposed a combinatorial polynomial-time algorithm for minimizing submodular functions. In fact, there are two algorithms in \cite{sat2001acspa}, the first algorithm runs in time bounded by a polynomial in the size of the underlying set and the length of the largest absolute function value, and the second one's running time is bounded by a polynomial in the size of the underlying set, independent of the function values. In our simulations, we find that the first algorithm is more efficient than the second one. This is due to the fact that the second algorithm requires more computation in each step in order to eliminate the effect of function values. So we apply the first algorithm in our problem. In addition, it needs to be noted that in order to use the first algorithm for our problem, we need to introduce a lower bound $\epsilon$ for the difference between the second minimum and the minimum value of the submodular function, and the lower bound also affects the number of steps of the algorithm.

We formally formulate the problem as
\begin{flalign}\label{equ_minpro}
\text{min}\,\,\,&f(S)=W\log(1+\tfrac{\sum_{i\in S}P_i}{N_0W})-\bm{R}(S)\\ \nonumber
\text{s.t. } &S\subseteq E=\{1,2,\ldots,N\}.
\end{flalign}

We define the \emph{submodular polyhedron} $P(f)$ and the \emph{base polyhedron} $B(f)$ as
\begin{align*}
& P(f)=\{\bm{x}|\bm{x}\in R^{E}, \forall S\subseteq E: \bm{x}(S)\leq f(S) \}, \\
& B(f)=\{\bm{x}|\bm{x}\in P(f), \bm{x}(E)= f(E) \},
\end{align*}
where $\bm{x}(S)=\sum_{i\in S}x(i)$. The \emph{submodular polyhedron} $P(f)$ is actually a polyhedron defined by $2^N-1$ inequalities, it is not a polymatroid since the function $f(S)$ does not satisfy the nondecreasing requirement. The vector in the \emph{base polyhedron} $B(f)$ is called a base, and an extreme point in $B(f)$ is called an extreme base.  The extreme bases can be calculated as follows

Let $L=\{v_1,v_2,\ldots,v_N\}$ be a linear ordering on the set $E=\{1,2,\ldots,N\}$, for any $i\in \{1,2,\ldots,N\}$, define $L(v_i)=\{v_1,\ldots,v_i\}$, then we can compute an \emph{extreme base} $\bm{x}\in B(f)$ associated with the ordering $L$ by
\begin{equation}\label{equ_extbase}
   x(v)=f(L(v))-f(L(v)\backslash \{v\}), \quad \forall v\in E.
\end{equation}

If $u$ immediately succeeds $v$ in a linear ordering, we can interchange $u, v$ and obtain a new ordering whose extreme base is given in the following lemma.

\emph{\textbf{Lemma 2 \cite{sat2001acspa}:}} \emph{The extreme base corresponding to the new linear ordering obtained by interchanging vertexes $v$ and $u$ ($u$ immediately succeeds $v$) in the ordering $L$ is
$$ 
  \bm{y}'=\bm{y}+\widetilde{c}(\bm{y},u,v)(\chi_u-\chi_v),
$$
where $\bm{y}$ is the extreme base corresponding to $L$ and
$$ 
  \widetilde{c}(\bm{y},u,v)=f(L(u)\backslash\{v\})-f(L(u))+y(v).
$$
 }

For any vector $\bm{x}\in R^{E}$, we use $\bm{x}^{+}$ to denote the vector with elements defined by $x^{+}(v)=\max\{0,x(v)\}$, and  $\bm{x}^{-}$ to denote the vector with elements defined by $x^{-}(v)=\min\{0,x(v)\}$. We use $\chi_u$ as the indicator vector such that $\chi (u)=1, \chi (v)=0, \forall v\neq u$.  It has been proven in \cite{sat2001acspa} that for a submodular function $f: 2^{E}\rightarrow R$,
\begin{equation}\label{equ_subequi}
  \max\{\bm{x}^{-}(E)|\bm{x}\in B(f)\}=\min\{f(S)|S\subseteq E \}.
\end{equation}

This algorithm seeks to find a base in the base polyhedron to maximize $\bm{x}^{-}(E)$. It does not maximize $\bm{x}^{-}(E)$ directly, but rather introduces another vector $\bm{z}=\bm{x}+\partial \phi$ and tries to maximize $\bm{z}^{-}$, where $\partial \phi$ is associated with a flow $\phi: E\times E\rightarrow R$ defined on the complete directed graph $G=(E,A)$, where $E$ is the vertices $\{1,2,\ldots,N\}$, $A$ is the arc set $E\times E$, and
\begin{equation}\label{equ_netflow}
  \partial \phi(u)=\sum\nolimits_{v\in E}\phi(u,v)-\sum\nolimits_{v\in E}\phi(v,u), \forall v\in E.
\end{equation}
$\partial \phi(u)$ is actually the net flow emanating from vertex $u$. We define the flow $\phi$ as $\delta-$feasible if it satisfies
$0\leq\phi(u,v)\leq \delta$, for all $u,v\in E$.

In order to guarantee that $\bm{x}$ is always in the base polyhedron, the algorithm maintains $\bm{x}$ as a convex combination of a set of extreme bases $\{\bm{y}_i, i\in I\}$ given in (\ref{equ_extbase}), where $I$ is an index set, i.e.
\begin{equation}\label{equ_convcom}
  \bm{x}=\sum\nolimits_{i\in I}\lambda_{i}\bm{y}_i.
\end{equation}
Besides, the algorithm also maintains the linear ordering $L_i$ associated with each extreme base $\bm{y}_i$.

Define the vertex sets $C=\{v|v\in E, z(v)\leq -\delta\}$ and $D=\{v|v\in E, z(v)\geq \delta\}$, in each $\delta-$scaling phase, the algorithm maintains a $\delta-$feasible flow $\phi$ and a subgraph $G^{\circ}=(V,A^{\circ})$, where the arc set $A^{\circ}=\{(u,v)|u,v\in E,u\neq v, \phi(u,v)=0\}$. The basic idea of this algorithm is to move flow from $C$ to $D$ along an $\delta-$augmenting path from the vertexes in $C$ to vertexes in $D$ in the subgraph $G^{\circ}$.

In cases where there is no such a $\delta-$augmenting path, the algorithm uses a procedure called Double-Exchange to adjust the flow. To illustrate the procedure of Double-Exchange, we first give the definition of the active triple: Let $B$ denote the vertexes currently reachable from $C$ in $G^{\circ}$. A triple $(i,u,v), i\in I, u, v\in E$,  is defined as an active triple when $u$ immediately succeeds $v$ in the ordering $L_{i}$ and $u\in B, v\in E\backslash B$. When applying Double-Exchange for an active triple, first compute the exchange capacity
$$\widetilde{c}(y_i,u,v)=f(L(u)\backslash\{v\})-f(L(u))+y_i(v),$$
then update $\bm{x}$ and $\phi$ as: $\bm{x}:=\bm{x}+\alpha(\chi_u-\chi_v),$ and $\phi(u,v):=\phi(u,v)-\alpha,$ where $\alpha=\min(\lambda_i\widetilde{c}(y_i,u,v),\phi(u,v))$, in this process, $\bm{z}=\bm{x}+\partial \phi$ remains unchanged. As a result, either $B$ remains unchanged or the vertex $v$ and other vertexes reachable from $v$ in $G^{\circ}$ are added to $B$. We also update the index set $I$, the extreme base set $\{\bm{y}_i, i\in I\}$, and the linear ordering set $\{L_i, i\in I\}$ accordingly. The process of Double-Exchange goes on until a $\delta-$augmenting path is found.

\begin{figure}
  \centering
  \includegraphics[width=0.48\textwidth]{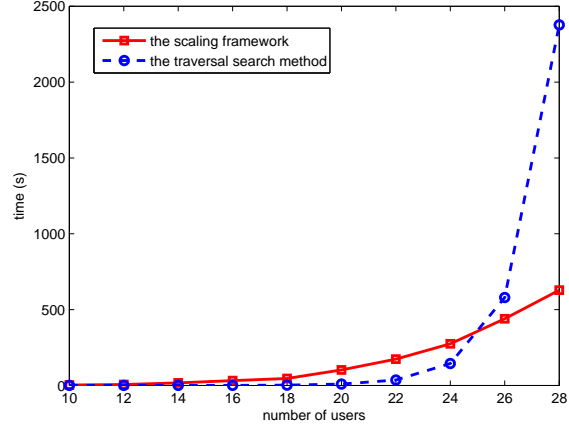}\\
  \caption{The running time of the scaling framework and traversal search method.}\label{fig_sfm_tsm}
\end{figure}

The algorithm terminates when there is neither a $\delta-$augmenting path nor an active triple. The formal description of the algorithm is referred to as the scaling framework in the following table.

\begin{tabular}{c @{  } l}
\hline
\multicolumn{2}{l}{\textbf{Algorithm 1: The Scaling Framework}} \\
\hline
1& \textbf{Initialization:} \\
2& \indent $L$ $\leftarrow$ a linear ordering on $E$ \\
3& \indent $\bm{x}$ $\leftarrow$ an extreme base in $B(f)$ generated by $L$ \\
4& \indent $\delta$ $\leftarrow$ $\min\{|\bm{x}^{-}(E)|,\bm{x}^{+}(E)\}/N^2$ \\
5& \indent $I$ $\leftarrow$ $\{k\}$, $y_k$ $\leftarrow$ $\bm{x}$, $\lambda_k$ $\leftarrow$ $1$, $L_k$ $\leftarrow$ $L$  \\
6& \indent $\phi$ $\leftarrow$ $\bm{0}$,  \\
7& \textbf{While} $\delta\geq 1/N^2$ \textbf{do} \\
8& \indent $C$ $\leftarrow$ $\{v|x(v)+\partial\phi(v)\leq -\delta \}$ \\
9& \indent $D$ $\leftarrow$ $\{v|x(v)+\partial\phi(v)\geq -\delta \}$ \\
10& \indent $B$ $\leftarrow$ the set of vertices reachable from $C$ in $G^{\circ}$ \\
11& \indent \textbf{While} $B\bigcap D\neq \emptyset$ or there is an active triple \textbf{do}  \\
12& \indent\indent \textbf{While} $B\bigcap D=\emptyset$ and there is an active triple\\
& \indent\indent \textbf{do} \\
13& \indent\indent\indent Apply \textbf{Double-Exchange} to an active triple \\
& \indent\indent\indent $(i,u,v)$ \\
14& \indent\indent\indent Update $B$.  \\
15& \indent\indent If $B\bigcap D\neq \emptyset$ then  \\
16& \indent\indent\indent    Augment flow $\phi$ along a $\delta$-augmenting path\\
&  \indent\indent\indent $P$ by setting $\phi(u,v):=\delta-\phi(v,u)$ \\
&  \indent\indent\indent and $\phi(v,u)=0$ for each arc $(u,v)$ in $P$.  \\
17& \indent\indent\indent   Update $G^{\circ}$, $C$, $D$, $B$.  \\
18& \indent\indent Apply \textbf{Reduce(x,I)}.  \\
19& \indent $\delta \leftarrow \delta/2$  \\
20& \indent $\phi \leftarrow \phi/2$  \\
21& \textbf{Return} $B$.  \\
22& \textbf{End.}  \\
\hline
\end{tabular}

The procedure of Double-Exchange is shown as follows.

\begin{tabular} {l @{  } l}
\hline
 \multicolumn{2}{l}{\textbf{Double-Exchange} $(i,u,v)$} \\
\hline
1& $\widetilde{c}(y_i,u,v)\leftarrow f(L_i(u)\backslash\{v\})-f(L_i(u))+y_i(v)$  \\
2& $\alpha\leftarrow \min\{\phi(u,v),\lambda_i\widetilde{c}(y_i,u,v)\}$  \\
3& $\bm{x}\leftarrow \bm{x}+\alpha (\bm{\chi}_{u}-\bm{\chi}_{v})$  \\
4& $\phi(u,v)\leftarrow \phi(u,v)-\alpha$   \\
5& \textbf{If} $\alpha <\lambda_i\widetilde{c}(y_i,u,v)$ \textbf{then} \\
6& \indent $k\leftarrow$ a new index  \\
7& \indent $I\leftarrow I\cup\{k\}$ \\
8& \indent $\lambda_{k}\leftarrow \lambda_i-\alpha/\widetilde{c}(y_i,u,v)$  \\
9& \indent $\lambda_i\leftarrow \alpha/\widetilde{c}(y_i,u,v)$  \\
10& \indent $\bm{y}_k\leftarrow \bm{y}_i$   \\
11& \indent $L_k\leftarrow L_i$  \\
12& $\bm{y}_i\leftarrow \bm{y}_i+\widetilde{c}(y_i,u,v)(\bm{\chi}_{u}-\bm{\chi}_{v})$ \\
13& Update $L_i$ by interchanging $u$ and $v$. \\
\hline
\end{tabular}

Let the set of extreme bases be $Y=\{\bm{y}_i, i\in I\}$, the corresponding set of coefficients be $\lambda=\{\lambda_i,i\in I\}$, then the procedure of $Reduce(\bm{x},I)$ in Algorithm 1 is

\begin{tabular} {l @{  } l}
\hline
 \multicolumn{2}{l}{\textbf{Reduce($\bm{x}$,I)} } \\
\hline
1&  \textbf{Initialization:}  \\
2& \indent $Y_{AI} \leftarrow \{\}$, $I_{AI} \leftarrow \{\}$, $\lambda_{AI} \leftarrow \{\}$  \\
3& \textbf{While} the set $Y$ is not empty \textbf{do}  \\
4& \indent take $\bm{y}_i$ from $Y$ to $Y_{AI}$, and the corresponding\\
 & \indent index $i$ to $I_{AI}$, coefficient $\lambda_i$ to $\lambda_{AI}$,    \\
5& \indent \textbf{if} there is a set of $\mu_j, j\in I_{AI}$ that is not \\
&  \indent identically $0$ and satisfies $\sum_{j\in I_{AI}}\mu_j \bm{y}_j=0$ \\
& \indent and $\sum_{j\in I_{AI}}\mu_j=0$ \textbf{then} \\
6& \indent\indent compute $\theta:= \min\{\lambda_j/\mu_j|\mu_j>0\}$ for $j\in I_{AI}$, \\
7& \indent\indent $\lambda_j:=\lambda_j-\theta \mu_j$ for $j\in I_{AI}$, \\
8& \indent\indent \textbf{for} $j\in I_{AI}$ \\
9&     \indent\indent\indent \textbf{if} $\lambda_j==0$ \textbf{then} \\
10&     \indent\indent\indent\indent delete $\bm{y}_j$ from $Y_{AI}$, $\lambda_j$ from $\lambda_{AI}$, and \\
& \indent\indent\indent\indent $j$ from $I_{AI}$.   \\
11& \textbf{return} $Y_{AI}$, $\lambda_{AI}$, $I_{AI}$.\\
\hline
\end{tabular}
\\
\\
It has been proven in \cite{sat2001acspa} that the steps required by the above algorithm is bounded by $O(N^5\log(M/\epsilon))$, where $M=\max\{|f(S)|,S\subseteq E\}$, and $\epsilon$ is the lower bound  for the difference between the second minimum and the minimum value of the submodular function.

We apply this algorithm in our problem, and do the simulations using Matlab. Since different people may write programs with different parameter settings, the running time may be different from person to person. Nevertheless, the trend is always the same, and we post our results in Fig. \ref{fig_sfm_tsm} to illustrate the advantage of the scaling framework over the traversal search method. It can be seen from this figure that when the number of users in the system is small, for instance, less than $25$, the traversal search method is more efficient, however, as the number of users increases, the time spent by using the traversal search method increases exponentially, while the scaling framework increases much slowly. When the number of users is larger than $25$, the scaling framework becomes more efficient.

\section{Feasible power allocation and optimization} \label{sec_poweralloc}

In cases where the multiaccess system does not meet the delay requirements of all users, we seek to find some ways such as power reallocation to make the system delay guarantee feasible.

To do so, we first assume that the sum power of the system is fixed. In this case, we find that as long as the sum power is larger than a threshold, it always makes the system meet the delay requirements of all users by adjusting the power allocation. This result is described by the following proposition:

\emph{\textbf{Proposition 2:}} \emph{If the sum power of the multiaccess system is larger than a threshold value, $(2^{\sum_{i=1}^{N}\frac{R_i}{W}}-1)N_0W$, where $R_i,i\in\{1,2,\ldots,N\}$ is the required service rate of user $i$, we can always make the system meet the delay requirements of all users by adjusting the transmission power of each user.}

The proof of  proposition 2 will be given after the following lemma.

\emph{\textbf{Lemma 3:}} \emph{The minimum required sum power of a multiaccess system is $\min\{\sum_{i=1}^{N}P_i\}=(2^{\sum_{i=1}^{N}\frac{R_i}{W}}-1)N_0W$, and we can always find a power allocation method to meet the delay requirements of all users with this minimum sum power.}
\begin{proof}
The multiaccess capacity region can be changed into a power region, for instance, the feasible power region of the multiaccess system of  (\ref{equ_capregion}) is
\begin{multline*}
  P_g(\bm{R})=\{\bm{P}:\bm{P}(S)\geq (2^{\sum_{i\in S}\frac{R_i}{W}}-1)N_0W,  \\
  \textrm{for}\,\,\textrm{ every} \quad S\subseteq\{1,\ldots,N\}\}.
\end{multline*}
As can be seen from the above region, the sum power should at least satisfy
\begin{equation} \label{equ_minsumpower}
\sum_{i=1}^{N}P_i\geq (2^{\sum_{i=1}^{N}\frac{R_i}{W}}-1)N_0W.
\end{equation}
So the minimum sum power should be no less than  $(2^{\sum_{i=1}^{N}\frac{R_i}{W}}-1)N_0W$. In order to prove lemma 2, we need to show that with this sum power, we can find a method to allocate the sum power to meet the rate requirements of all users.

One possible power allocation method is given by the following equation
\begin{equation}\label{equ_powalloc}
  P_j=\frac{(2^{\frac{R_j}{W}}-1)(2^{\sum\nolimits_{i=1}^{N}\frac{R_i}{W}}-1)N_0W}{\sum\nolimits_{i=1}^{N}(2^{\frac{R_i}{W}}-1)}.
\end{equation}
To show that the above power allocation method is feasible, we need to prove that for all $S\subseteq E=\{1,2,\ldots,N\}$, there is $\bm{P}(S)\geq (2^{\sum_{j\in S}\frac{R_j}{W}}-1)N_0W$, substituting Eqn. (\ref{equ_powalloc}) in the inequality, we have
\begin{equation*}
  \frac{\sum\nolimits_{j\in S}(2^{\frac{R_j}{W}}-1)(2^{\sum\nolimits_{i=1}^{N}\frac{R_i}{W}}-1)N_0W}{\sum\nolimits_{i=1}^{N}(2^{\frac{R_i}{W}}-1)}\geq (2^{\sum\nolimits_{j\in S}\frac{R_j}{W}}-1)N_0W
\end{equation*}
or equivalently,
\begin{equation}\label{equ_ieqpro}
  \frac{\sum\nolimits_{j\in S}(2^{\frac{R_j}{W}}-1)}{2^{\sum\nolimits_{j\in S}\frac{R_j}{W}}-1}\geq \frac{\sum\nolimits_{i=1}^{N}(2^{\frac{R_i}{W}}-1)}{2^{\sum\nolimits_{i=1}^{N}\frac{R_i}{W}}-1}.
\end{equation}

Define the set function $g(S)=\frac{\sum\nolimits_{j\in S}(2^{\frac{R_j}{W}}-1)}{2^{\sum\nolimits_{j\in S}\frac{R_j}{W}}-1}$, we  show that $g(S)$ arrives its minimum when the set $S=E$.

Take a set $S=\{1,2,\ldots,k\}$, and $S'=S\cup\{k+1\}$, if $g(S)\geq g(S')$, i.e.
\begin{equation}\label{equ_ieqmono}
  \frac{\sum\nolimits_{j\in S}(2^{\frac{R_j}{W}}-1)}{2^{\sum\nolimits_{j\in S}\frac{R_j}{W}}-1}\geq \frac{\sum\nolimits_{j\in S'}(2^{\frac{R_j}{W}}-1)}{2^{\sum\nolimits_{j\in S'}\frac{R_j}{W}}-1},
\end{equation}
then Inequality (\ref{equ_ieqpro}) will hold.

Inequality (\ref{equ_ieqmono}) is equivalent to

\begin{equation} \label{iequ_subtr}
\begin{split}
&(\sum\nolimits_{j\in S}(2^{\frac{R_j}{W}}-1))(2^{\sum\nolimits_{j\in  S'}\frac{R_j}{W}}-1)-(\sum\nolimits_{j\in S'}(2^{\frac{R_j}{W}}-1)) \\
& (2^{\sum\nolimits_{j\in S}\frac{R_j}{W}}-1) \\
=& ((\sum\nolimits_{j\in S}(2^{\frac{R_j}{W}}-1))2^{\sum\nolimits_{j\in S}\frac{R_j}{W}}-2^{\sum\nolimits_{j\in S}\frac{R_j}{W}}+1)\\
&(2^{\frac{R_{k+1}}{W}}-1)\\
\geq &  0,
\end{split}
\end{equation}
since $R_i\geq 0$, $2^{\frac{R_{i}}{W}}-1\geq 0$, so we only need to prove
$(\sum\nolimits_{j\in S}(2^{\frac{R_j}{W}}-1))2^{\sum\nolimits_{j\in S}\frac{R_j}{W}}-2^{\sum\nolimits_{j\in S}\frac{R_j}{W}}+1\geq 0$.
This inequality can be proven by induction.

First, it is easy to see that when $S=\{1\}$, the inequality holds since
\begin{equation*}
 (2^{\frac{R_1}{W}}-1)2^{\frac{R_1}{W}}-2^{\frac{R_1}{W}}+1\\
 =(2^{\frac{R_1}{W}}-1)^2\geq 0,
\end{equation*}

second, assume that when $S=\{1,2,\ldots,k\}$, the inequality holds, we need to prove it also holds when $S'=S\cup\{k+1\}$,

\begin{equation} \label{equ_interme}
\begin{split}
&(\sum\nolimits_{j\in S'}(2^{\frac{R_j}{W}}-1))2^{\sum\nolimits_{j\in S'}\frac{R_j}{W}}-2^{\sum\nolimits_{j\in S'}\frac{R_j}{W}}+1 \\
=&(\sum\nolimits_{j\in S}(2^{\frac{R_j}{W}}-1))2^{\sum\nolimits_{j\in S}\frac{R_j}{W}}2^{\frac{R_{k+1}}{W}}+(2^{\frac{R_{k+1}}{W}}-1)2^{\sum\nolimits_{j\in S'}\frac{R_j}{W}} \\
&-2^{\sum\nolimits_{j\in S'}\frac{R_j}{W}}+1,
\end{split}
\end{equation}
since when $S=\{1,2,\ldots,k\}$, we have
$$(\sum\nolimits_{j\in S}(2^{\frac{R_j}{W}}-1))2^{\sum\nolimits_{j\in S}\frac{R_j}{W}}\geq 2^{\sum\nolimits_{j\in S}\frac{R_j}{W}}-1,$$
substituting it in Eqn. (\ref{equ_interme}), we can get
\begin{equation}
\begin{split}
  & (2^{\sum\nolimits_{j\in S}\frac{R_j}{W}}-1)2^{\frac{R_{k+1}}{W}}+(2^{\frac{R_{k+1}}{W}}-1)2^{\sum\nolimits_{j\in S'}\frac{R_j}{W}}-\\
  & 2^{\sum\nolimits_{j\in S'}\frac{R_j}{W}}+1 \\
  =& (2^{\frac{R_{k+1}}{W}}-1)(2^{\sum\nolimits_{j\in S'}\frac{R_j}{W}}-1)\geq 0.
\end{split}
\end{equation}

Therefore, for all $S\subseteq E=\{1,2,\ldots,N\}$, $$(\sum\nolimits_{j\in S}(2^{\frac{R_j}{W}}-1))2^{\sum\nolimits_{j\in S}\frac{R_j}{W}}-2^{\sum\nolimits_{j\in S}\frac{R_j}{W}}+1\geq 0.$$ Then Inequalities (\ref{iequ_subtr}), (\ref{equ_ieqmono}) and (\ref{equ_ieqpro}) hold in turn, and the power allocation method of Eqn. (\ref{equ_powalloc}) is feasible.
\end{proof}

Next, we will prove Proposition 2:
\begin{proof}
If the sum power of the system is $P_{sum}\geq (2^{\sum_{i=1}^{N}\frac{R_i}{W}}-1)N_0W$, we can give a power allocation method as
\begin{equation}\label{equ_powerallo_fixed}
P_j=\frac{(2^{\frac{R_j}{W}}-1)P_{sum}}{\sum\nolimits_{i=1}^{N}(2^{\frac{R_i}{W}}-1)}.
\end{equation}

From the proof of lemma 2, it is easy to see that this power allocation method is feasible, and on the other hand, the sum power of this system remains the same.
\end{proof}

\textbf{\emph{Remark:}} \emph{A power allocation method is optimal when the corresponding sum power is the minimum.}

From the above discussions, we see that Eqn. (\ref{equ_powalloc}) is optimal.

\begin{figure}
  \centering
  \includegraphics[width=0.48\textwidth]{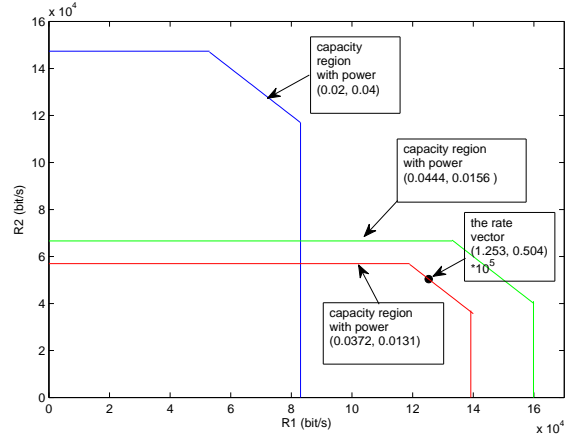}\\
  \caption{Power allocation for two users to meet the delay requirements. }\label{fig_poweralloc}
\end{figure}

To illustrate this, we first consider a $2$ users case, in which the average packet arrival rates are $\lambda_1=800\, bit/s,\, \lambda_2=600\, bit/s$, the corresponding delay requirements are $\tau_1=20\, \mu s,\, \tau_2=8\, \mu s$, and the original transmission powers of the two users are $P_1=20\, mW,\, P_2=40\, mW$. The bandwidth is $200 kHz$, and the power spectrum density of noise is $3\times 10^{-7} W/Hz$.  By Eqn. (\ref{equ_rsrv}), we find the required service rate vector as $\bm{R}=(1.253, 0.504)\times 10^5\, bit/s$. As can be seen from Fig. \ref{fig_poweralloc}, the required rate vector lies outside the capacity region defined by the original power allocation (the region defined by the blue line ), therefore, the multiaccess system cannot meet the delay requirements of the two users.

Next, let us consider power reallocation. From Eqn. (\ref{equ_powerallo_fixed}), we obtain a feasible power allocation method with the same sum power, which is $P_1=44.4\, mW, P_2=15.6\, mW$. By reallocating the sum power, we include the rate vector in the new capacity region defined by the green line. Furthermore, let us consider the optimal power allocation mode. By Inequatlity (\ref{equ_minsumpower}), the minimum sum power is $50.3\, mW$, and the optimal power allocation method is $P_1=37.2\, mW,\, P_2=13.1\, mW$. The corresponding capacity region is marked by the red line in Fig. \ref{fig_poweralloc}, it includes the rate vector on one edge and therefore can satisfy the delay requirements as well.

Furthermore, we provide an example of 3 users. The average packet arrival rates are $\lambda_1=919.54\, bit/s,\, \lambda_2=642\, bit/s,\, \lambda_3=105.32\, bit/s$, the corresponding delay requirements are $\tau_1=23\, \mu s,\, \tau_2=29.9\, \mu s,\, \tau_3=6.83\, \mu s$, and the original transmission powers of the three users are $P_1=0.5561\, W,\, P_2=0.0050\, W,\, P_3=0.4948\, W$. The bandwidth is $200 kHz$, and the power spectrum density of noise is $3\times 10^{-7} W/Hz$. The required service rate vector can be calculated from Eqn. (\ref{equ_rsrv}) as $\bm{R}=(0.4394, 0.3377, 1.4647)\times 10^5\, bit/s$.

\begin{figure}
  \centering
  \includegraphics[width=0.48\textwidth]{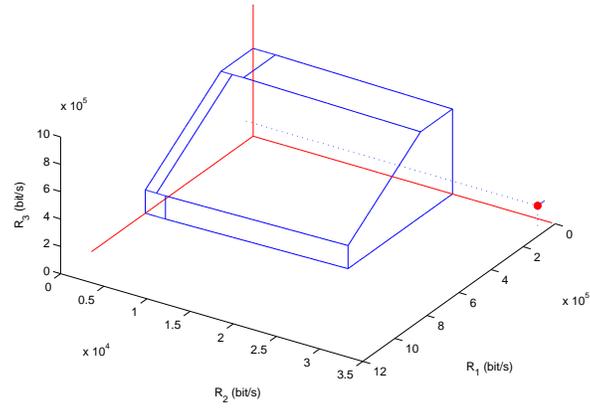}\\
  \caption{Original power allocation for three users. }\label{fig_original_3users}
\end{figure}

As can be seen from Fig. \ref{fig_original_3users}, the required service rate lies outside of the capacity region, thus, the system cannot satisfy the delay requirements of the 3 users.

\begin{figure}
  \centering
  \includegraphics[width=0.48\textwidth]{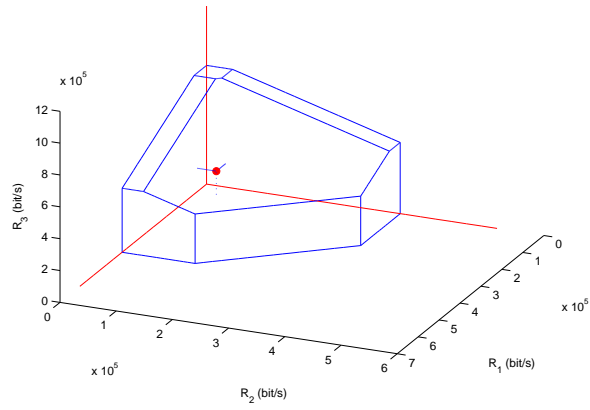}\\
  \caption{A feasible power allocation for three users. }\label{fig_feasible_3users}
\end{figure}

We then consider reallocating the sum power according to Eqn. (\ref{equ_powerallo_fixed}) to provide a feasible solution. The reallocated powers of the three users are $P_1=0.1828\, W,\, P_2=0.1380\, W,\, P_3=0.7351\, W$. The new capacity region includes the rate vector inside, as shown in Fig. \ref{fig_feasible_3users}.

\begin{figure}
  \centering
  \includegraphics[width=0.48\textwidth]{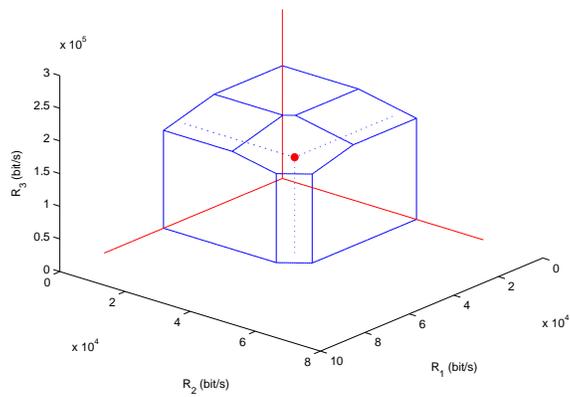}\\
  \caption{An optimal power allocation for three users. }\label{fig_optimal_3users}
\end{figure}

The optimal power allocation is obtained from Eqn. (\ref{equ_powalloc}) as $P_1=0.0122\, W,\, P_2=0.0092\, W,\, P_3=0.0491\, W$. The rate vector lies on one facet of the capacity region, as shown in Fig. \ref{fig_optimal_3users}.

\section{Conclusion} \label{sec_conclusion}

In this paper, we combined queueing theory and information theory to analyze the delay requirements guarantee problem in a multiaccess system. In particular, we divided the problem into two separate layers--the multaccess layer and the physical layer. In the multiaccess layer, we proved that in order to minimize the service rate of each user with a minimum delay requirement, it is better to allocate a fixed rate rather than to allocate the rate according to some distribution. In the physical layer, we discussed how to determine whether the rate vector lies in the multiaccess region within polynomial-time steps, and provided a polynomial-time algorithm based on submodular function minimization. In addition, we discussed how to adjust the transmission power of each user to make the system delay feasible provided that the sum power is larger than a minimum one.

It needs to be noted that, in this work, we only considered the Gaussian additive multiaccess channel, and fading is not involved, our future work would take the effect of fading into account, and discuss how to allocate the sum power in a fading environment. Besides, the packet length is assumed fixed and normalized to $1$ in this paper, we would like to extend this work to a situation with variable packet lengths. Finally,  in cases where the multiaccess system can meet the different delay requirements, how to allocate the rate in order to optimize the system performance is another possible direction for our future research.

\section{Acknowledgement}

This work was partly supported by the China Major State Basic Research Development Program (973 Program) No.2012CB316100(2), National Natural Science Foundation of China(NSFC) No. 61201203 and No.61171064, and the State Key
Laboratory of Rail Traffic Control and Safety,
Beijing Jiaotong University under Grant No.
RCS2012ZT008.

\bibliographystyle{ieeetr}
\bibliography{Delay_Feasible}

\begin{thebibliography}{10}

\bibitem{yeh2002doramac}
E.~Yeh, ``Delay-optimal rate allocation in multiaccess communications: a
  cross-layer view,'' in {\em Multimedia Signal Processing, 2002 IEEE Workshop
  on}, pp.~404--407, Dec. 2002.

\bibitem{mandal2012drpma}
P.~Mandal and S.~De, ``A new dynamic reservation protocol for many-to-one
  multi-access with long propagation delay,'' in {\em Vehicular Technology
  Conference (VTC Fall), 2012 IEEE}, pp.~1--5, Sep. 2012.

\bibitem{ehsan2001dotpwmc}
N.~Ehsan and T.~Javidi, ``Delay optimal transmission policy in a wireless
  multiaccess channel,'' {\em Information Theory, IEEE Transactions on},
  vol.~54, pp.~3745--3751, Aug. 2001.

\bibitem{jing2010dmtfapcmac}
J.~Yang and U.~Sennur, ``Delay-minimal transmission for average power
  constrained multi-access communications,'' {\em Wireless Communications, IEEE
  Transactions on}, vol.~9, pp.~2754 -- 2767, Sep. 2010.

\bibitem{telatar1995qtitm}
I.~E. Telatar and R.~G. Gallager, ``Combining queueing theory with information
  theory for multiaccess,'' {\em Selected Areas in Communications, IEEE Journal
  on}, vol.~13, pp.~963 -- 969, Aug. 1995.

\bibitem{wolff1989smattq}
R.~W. Wolff, {\em Stochastic Modeling and the Theory of Queues}.
\newblock Prentice-Hall International, Inc, 1989.

\bibitem{tse1998multfc}
D.~N.~C. Tse and S.~V. Hanly, ``Multiaccess fading channels-part i: polymatroid
  structure,optimal resource allocation and throughput capacities,'' {\em
  Information Theory, IEEE Transactions on}, vol.~44, pp.~2796--2815, Nov.
  1998.

\bibitem{sat2001acspa}
S.~Iwata, L.~Fleischer, and S.~Fujishige, ``A combinatorial strongly polynomial
  algorithm for minimizing submodular functions,'' {\em Journal of ACM},
  vol.~48, pp.~761--777, Jul. 2001.

\bibitem{gro1981temac}
M.~GR\"{O}TSCHEL, L.~LOV\'{A}SZ, and A.~SCHRIJVER, ``The ellipsoid method and
  its consequences in combinatorial optimization,'' {\em Combinatorica},
  pp.~169--197, 1981.

\bibitem{whc1984tmimp}
W.~H. Cunningham, ``Testing membership in matroid polyhedra,'' {\em Journal of
  Combinatorial Theory, Series B 36}, pp.~161--188, 1984.

\end{thebibliography}
\end{document}